\documentclass[prb,twocolumn,amsmath,showpacs]{revtex4}

\usepackage{graphicx}

\begin{document}
\input epsf

\title {Ambivalence of the anisotropy of the vortex lattice 
in an anisotropic type-II superconductor.}

\author {I. L. Landau$^{1,2}$, A. V. Sologubenko$^{1}$, H. R. Ott$^{1}$}
\affiliation{$^{1}$Laboratorium f\"ur Festk\"orperphysik, ETH 
H\"onggerberg, CH-8093 Z\"urich, Switzerland}
\affiliation{$^{2}$Kapitza Institute for Physical Problems, 117334 Moscow, 
Russia}

\date{\today}

\begin{abstract}
We present a geometry-based discussion of possible vortex configurations 
in the mixed state of anisotropic type-II superconductors. It is shown 
that, if energy considerations assign six nearest neighbors to each 
vortex, two distinct modifications of the vortex lattice are possible. 
It is expected that certain conditions lead to a first order phase 
transition from one modification of the vortex lattice to the other upon 
varying the external magnetic field. 
\end{abstract}

\maketitle

In anisotropic type-II superconductors the upper critical field $H_{c2}$, 
the magnetic-field penetration depth $\lambda$, and the coherence length 
$\xi$ depend on the direction with respect to crystallographic axes of 
the material and these dependencies may schematically be represented by 
ellipsoids. We consider the case where the mixed state is created by an 
external magnetic field  directed along one of the principal axes of 
these ellipsoids, say the $y$-axis. To keep the discussion simple, the 
above mentioned quantities are assumed to be isotropic in the $xy$-plane 
and $H_{c2}^{(xy)} > H_{c2}^{(z)}$. We also assume that in the fully 
isotropic case, the energetically most favorable arrangement assigns six 
nearest neighbors to each vortex (triangular lattice). 

We use the commonly accepted notation where the anisotropy is expressed 
as
\begin{equation}
\gamma = \frac{H_{c2}^{(xy)}}{H_{c2}^{(z)}} = 
\frac{\xi^{(xy)}}{\xi^{(z)}} = \frac{\lambda^{(z)}}{\lambda^{(xy)}}
\end{equation}
In this case, $\lambda^{(z)}$ is the screening length corresponding to 
screening currents oriented along the $z$-axis.\cite{tink} The 
anisotropies of $\lambda$ and $\xi$ in the plane perpendicular to the 
magnetic field, i.e., the $xz$-plane, may both be represented by 
ellipses with the same eccentricities and elongated along the $x$-axis, 
as is shown in Fig. 1. 
\begin{figure}[h]
 \begin{center}
  \epsfxsize=0.95\columnwidth \epsfbox {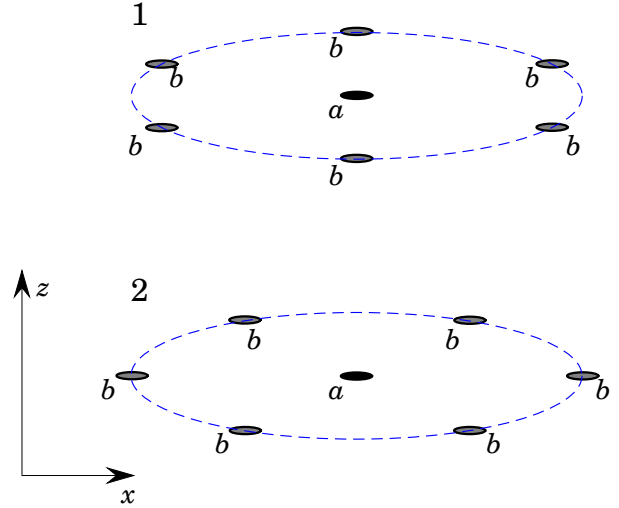}
  \caption{Two possible configurations of vortices for $H$ oriented 
           along $y$-axis and assuming $\gamma = 3.5$. The small 
           ellipses represent single vortices, their shape indicates 
           the anisotropy of $\xi$.}
 \end{center}
\end{figure}

In magnetic fields well below $H_{c2}$, it is the magnetic interaction 
between the vortices that provides the main contribution to the free 
energy of the mixed state. This means that the distances between one 
chosen vortex line and its six nearest neighbors should all be equal in 
units of $\lambda$, i.e., the nearest neighbors are situated along 
ellipses that reflect the anisotropy of $\lambda$. As was shown in Ref. 
\onlinecite{kogan}, if $H \ll H_{c2}$ is oriented along the $xy$-plane, 
all such vortex arrangements have the same free energy, independent of 
the orientation of the triangular unit cell in the $xz$-plane. It is 
well known, however, that part of the free energy arises from the 
interaction of Abrikosov vortices with the crystal lattice. \cite{ess} 
This is why we argue that only the two most symmetrical configurations, 
shown in Fig. 1, are of practical interest. These lattices differ by their 
orientations with respect to the $x$- and the $z$-axis. Although 
lattice 1 and lattice 2 appear as significantly different, both of them 
correspond to the same average vortex density and the distances between 
the central vortex line $(a)$ and its nearest neighbors $(b)$ are also 
the same in units of $\lambda$.
\begin{figure}[h]
 \begin{center}
  \epsfxsize=0.95\columnwidth \epsfbox {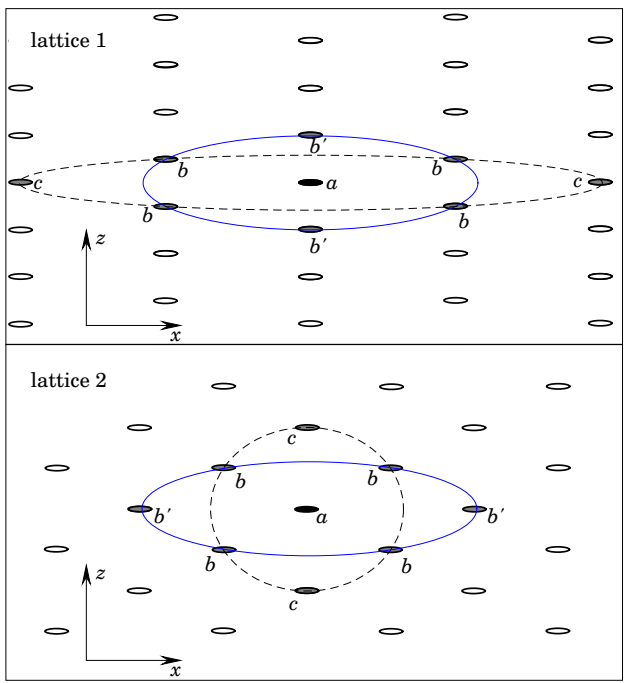}
  \caption{Two possible vortex lattices. Although both these lattices 
           are drawn for an anisotropy $\gamma = 3.5$ (solid-line 
           ellipses), they are equally consistent with anisotropies 
           that differ from $\gamma$ by a factor of 3, as is 
           emphasized by the dashed-line ellipses (see text for 
           details).}
 \end{center}
\end{figure}

The vortex lattice arrangements that follow from the two configurations 
shown in Fig. 1 are displayed in Fig. 2. While the vortex lattice 1 is 
indeed strongly anisotropic and its geometrical appearance is  as it 
might be expected intuitively, lattice 2, on the other hand, looks much 
less anisotropic. In real space the nearest neighbors of the vortex line $a$ 
are the vortices $b$ and $c$. These vortices form a pattern with a close 
to triangular symmetry, which is emphasized by the dashed-line ellipse 
in the bottom panel of Fig. 2. At the same time, if we consider the 
magnetic interaction between the vortices, the distances must be measured 
in units of $\lambda$ and in this renormalized space, the nearest neighbors 
of vortex $a$ are the vortices $b$ and $b'$, as is illustrated by the 
solid-line ellipse in the bottom panel of Fig. 2. Thus, although both 
vortex lattices shown in Fig. 2 correspond to the same anisotropy of 
$\lambda$, the anisotropies of the corresponding vortex lattices in real 
space are quite different. The real space anisotropy of vortex lattice 
2 may be characterized by the eccentricity $\epsilon^{(2)}$ of the ellipse 
drawn through the nearest neighbors of one chosen vortex line (dashed-line 
ellipse in Fig. 2). A simple calculation reveals that if for the vortex 
lattice 1, $\epsilon^{(1)} = \gamma$, the eccentricity for the lattice 
2 is $\epsilon^{(2)} = \gamma/3$. This means that if, e.g., the original 
anisotropy is $\gamma = 3$, the vortex lattice 2 is perfectly isotropic 
in real space.

Although the two lattices shown in Fig. 2 are quite different in real 
space, in both cases the distances between the nearest neighbors in the 
renormalized space are the same in units of $\lambda$. Furthermore, if 
Eq. (1) is exact, i.e., in frameworks of the applicability of the 
Ginzburg-Landau theory, the distances between the vortices are also equal 
in units of $\xi$. This makes the free energies of the two configurations 
identical not only in low magnetic field limit but in the entire magnetic 
field range $H_{c1} \le H \le H_{c2}$. In this case, minor contributions 
to the free energy depending, e.g., on the symmetry of the crystal 
lattice, may change the energy balance in favor of one or the other 
configuration. Our simple arguments cannot serve to answer the 
question, which of the two considered vortex lattices is stable. Depending 
on the particular chosen superconductor, any of the two may energetically 
be favored.

The situation is different if the Ginzburg-Landau theory is not 
quantitatively applicable, for instance, at temperatures well below 
$T_c$. In such cases the anisotropies $\gamma_{\xi} = \xi^{(xy)} / 
\xi^{(z)}$ and $\gamma_{\lambda} = \lambda^{(z)} / \lambda^{(xy)}$ can 
be different. This circumstance changes the energy balance between the 
two considered vortex configurations in higher magnetic fields where 
overlapping of vortex cores has to be taken into account. It is easy 
to check that if $\gamma_{\xi} > \gamma_{\lambda}$, lattice 1 has a 
lower free energy and vise versa. In the case of $\gamma_{\xi} \ne 
\gamma_{\lambda}$, it is possible that the energy balance between the two 
vortex lattices depends on the applied magnetic field. If this is indeed 
the case, we expect a first order phase transition from one vortex 
configuration to the other with increasing magnetic field. 

The main result of our consideration is that the symmetry of the vortex 
lattice does not provide unambiguous information about the anisotropy 
$\gamma$ of the magnetic field penetration depth. It turns out that 
each vortex configuration is consistent with two unequal values of 
$\gamma$, differing by a factor of 3. For instance, vortex lattice 1 
equally well corresponds to $\gamma = 3.5$ and $\gamma = 10.5$, while 
lattice 2 reflects the anisotropies of 3.5 and 1.17, as it is shown 
by the solid and dashed-line ellipses in Fig. 2. 

Recognizing the fact that the anisotropy of the vortex lattice in 
real space may be quite different from that of $\lambda$ is rather 
important for the correct interpretation of experimental observations. 
With this in mind we recall the recent experimental observation of the 
vortex lattice in MgB$_2$, invoked by magnetic fields oriented along the 
$ab$-planes of the hexagonal crystal lattice. \cite{ensk} A hexagonal 
vortex lattice corresponding to an eccentricity $\epsilon = 1.19$ was 
observed in these experiments. This value of $\epsilon$ is much smaller 
than the corresponding anisotropy $\gamma$ of $H_{c2}$. Although some 
theoretical justifications for the anisotropy of the magnetic field 
penetration depth $\gamma_{\lambda} = \lambda^{(z)}/\lambda^{(xy)}$ to 
be smaller than $\gamma_{H} = H_{c2}^{(xy)}/H_{c2}^{(z)} = 
\xi^{(xy)}/\xi^{(z)}$ were offered in Ref. \onlinecite{ensk}, we argue 
that the observation of Ref. \onlinecite{ensk} most likely reflects 
a $\lambda$-anisotropy $\gamma_{\lambda} = 3\epsilon 
\approx 3.6$ (see the bottom panel of Fig. 2). This latter value is in 
fair agreement with the observed anisotropy of $H_{c2}$.\cite{ani}

In conclusion, we showed that for each value of the anisotropy of the 
magnetic field penetration depth $\gamma_{\lambda}$ there are two 
possible arrangements of the hexagonal vortex lattice with 
corresponding eccentricities $\epsilon^{(1)} = \gamma_{\lambda}$ and 
$\epsilon^{(2)} = \gamma_{\lambda}/3$. This is why an unambiguous 
determination of $\gamma_{\lambda}$ from experimental observations of 
the vortex lattice is only possible if an approximate value of 
$\gamma_{\lambda}$ is {\it a priori} known.

\end{document}